\documentclass[conference]{IEEEtran}
\IEEEoverridecommandlockouts
\usepackage{cite}
\usepackage{amsmath,amssymb,amsfonts}
\usepackage{algorithmic}
\usepackage{graphicx}
\usepackage{textcomp}
\usepackage{xcolor}
\def\BibTeX{{\rm B\kern-.05em{\sc i\kern-.025em b}\kern-.08em
T\kern-.1667em\lower.7ex\hbox{E}\kern-.125emX}}
\begin{document}
\title{Digital Twin-Native AI-Driven Service Architecture for Industrial Networks\\
}

\author{
\IEEEauthorblockN{
Kübra Duran\IEEEauthorrefmark{1}\IEEEauthorrefmark{2},
Matthew Broadbent\IEEEauthorrefmark{1},
Gökhan Yurdakul\IEEEauthorrefmark{3}, and 
Berk Canberk\IEEEauthorrefmark{1}}\\
\IEEEauthorblockA{\IEEEauthorrefmark{1}School of Computing, Engineering and The Build Environment, Edinburgh Napier University, United Kingdom \\
\IEEEauthorrefmark{2}Computer Engineering Department, Istanbul Technical University, Turkey \\
\IEEEauthorrefmark{3}BTS Group, Istanbul, Turkey \\
Email: \{K.Duran, M.Broadbent, B.Canberk\}@napier.ac.uk, \\durank18@itu.edu.tr, yurdakulg@btsgrp.com}
}

\maketitle

\begin{abstract}
The dramatic increase in the connectivity demand results in an excessive amount of Internet of Things (IoT) sensors. To meet the management needs of these large-scale networks, such as accurate monitoring and learning capabilities, Digital Twin (DT) is the key enabler. However, current attempts regarding DT implementations remain insufficient due to the perpetual connectivity requirements of IoT networks. Furthermore, the sensor data streaming in IoT networks cause higher processing time than traditional methods. In addition to these, the current intelligent mechanisms cannot perform well due to the spatiotemporal changes in the implemented IoT network scenario. To handle these challenges, we propose a DT-native AI-driven service architecture in support of the concept of IoT networks. Within the proposed DT-native architecture, we implement a TCP-based data flow pipeline and a Reinforcement Learning (RL)-based learner model. We apply the proposed architecture to one of the broad concepts of IoT networks, the Internet of Vehicles (IoV). We measure the efficiency of our proposed architecture and note ~30\%  processing time-saving thanks to the TCP-based data flow pipeline. Moreover, we test the performance of the learner model by applying several learning rate combinations for actor and critic networks and highlight the most successive model.
\end{abstract}


\section{Introduction}
Over the recent years, Digital Twin (DT) has been raised as one of the most stimulating technological developments to shape tomorrow’s industrial environments\cite{book1}. Especially for large-scale Internet of Things (IoT) networks with an excessive amount of IoT sensors, the need for monitoring and efficient management has increased the importance of DT integration. Moreover, Internet of Vehicles (IoV) networks enable data exchange and intelligent collaboration between various objects, such as vehicles, environmental sensors throughout roads, and traffic light sensors. As is one of the broader concepts of IoT, the similar needs and the DT integration trend apply to IoV networks in the monitoring and management processes. 

Here, several challenges need to be addressed to implement a DT integration considering the broad concept of IoT networks. Such a design should address the following challenges: 

\begin{itemize}
    \item \textit{Perpetual Connectivity:} The conventional one-way interaction between sensors and their implemented models is insufficient to serve seamless integration and, thus, accurate monitoring due to the lack of the feedback interface. Therefore, the communication links between the sensors and their virtual models should be maintained in both directions continuously.
    \item \textit{Sensor Data Streaming:} The DT representation of physical world sensors provides exact monitoring to keep track of the congestion occurrence. Therefore, gathering the sensor's information plays a key role in the design of a DT-based service architecture. At this point, the question 
    \textit{“How much granularity and processing is required?”} should be decided according to the application of the IoV network to design an efficient data streaming approach.
    \item \textit{Depth of Data Insights:} Although there are concrete examples of Artificial Intelligence/Machine Learning (AI/ML) applications in the management of industrial networks, they become insufficient against spatiotemporal changes in the application environment. Therefore, enhanced learning models with high stability should be developed to maintain a deep understanding and precise behaviour modelling of the IoT environment.
\end{itemize}

\begin{figure*}[htbp]
\centerline{\includegraphics[width=\linewidth]{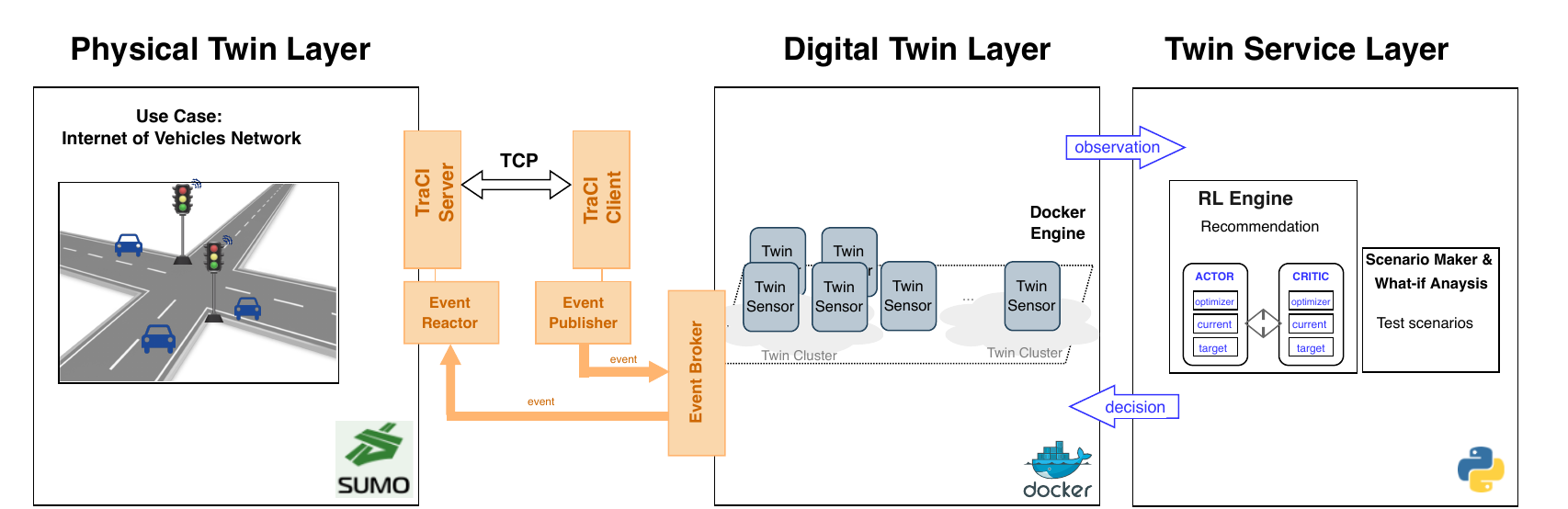}}
\caption{Proposed Digital Twin-native service architecture for IoV use case.}
\label{one}
\end{figure*}

\section{Related Works}

Most of the current literature for IoV networks focuses on vehicle routing strategies considering vehicle information rather than environmental information. For instance, \cite{l2} uses edge and cloud computing to reduce traffic congestion by proposing optimized routes for each connected vehicle. Likewise, \cite{l3} proposes a route recommendation framework to produce the shortest path for each source-destination pair promptly. Moreover, \cite{l4} uses a deep reinforcement learning algorithm to manage traffic light timings for road management systems. However, the whole IoV network could not be presented realistically in these studies due to the lack of considering a divergent set of environmental entity information.
On the other hand, despite the increasing trend in deploying DT in road management systems, the enhanced DT modelling challenge still needs to be addressed. None of the studies on transportation systems maintains a specific data flow scheme to handle high-quality representation needs by preserving the implementation needs. For instance, \cite{l5} studies DT modelling and points out a data representation framework by holding relation information; however, it does not implement the proposed model on road transportation networks. Also, a DT assisted Mobile Edge Computing (MEC) framework for Industrial IoT (IIoT) networks is proposed in \cite{MEC-2} to decrease the end-to-end latency. Furthermore, \cite{l6} establishes a traffic infrastructure's efficiency assessment model based on DTs with traditional database models, resulting in high processing time. Similarly, \cite{l7} introduces a learning approach to investigate the impact of taxi trips from an environmental issues perspective. Although these studies perform DT platform development efforts, they result in high querying and processing time with the traditional modelling approaches. Regarding AI-driven service provisioning, \cite{ccnc} and \cite{akeli} apply ML algorithms in order to come up with a stable network architecture for core networks. Moreover, there is the application of AI methods that focuses on the estimation of emission levels. For example, \cite{l8}, and \cite{l9} propose an emission regulatory framework for transportation networks by estimating the emission levels. In addition, \cite{l10} conducts emission prediction for vehicles using data generated by in-vehicle sensors by utilizing  LSTM models. Moreover, \cite{l11} predicts exhaust emissions by using multiple 1D-Convolutional Neural Networks (CNN) and Artificial Neural Networks (ANNs). Also, \cite{tgcn} estimates the discovery states in a core network architecture by proposing a DT-based approach. Although the results of these works demonstrate that DT provides accurate estimation results, there is no recommendation scheme implemented by taking into consideration the particular event definitions for sensors.

Recent advancements in DT for IoV networks have been described above. Nevertheless, none of these studies deals with the perpetual connectivity, sensor data streaming, and qualified data insights challenges, as listed in the introduction section. Thus these challenges remain unresolved. For this reason, this study is situated around the research question, \textit{“How to serve (i) a perpetual communication link between a physical entity and its virtual model to ensure seamless integration, (ii) an efficient sensor data streaming method with the target of preserving processing time, and (iii) an intelligent service model with the capability of accurate behaviour modelling in IoT networks scenario?”}. To address this, we propose a DT-native AI-driven service architecture consisting of a TCP-based data flow pipeline and a Reinforcement Learning (RL)-based learner model. We summarize the contributions of this study below:  
\begin{itemize}
    \item We implement an Internet of Vehicles (IoV) network scenario as the physical network and present its realistic representation by enabling the connection via the Traffic Control Interface (TraCI).
    \item We introduce a novel DT-native TCP-based data flow pipeline between the physical and digital twin layer of the IoV network to serve sensor data streaming in an event-based manner and trigger the reverse data flow, in other words, feedback interface, if required.
    \item We propose a learner model with Deep Deterministic Policy Gradient (DDPG) algorithm to experience the IoV environment by learning the behaviours and recommend optimal rules to avoid congestion occurrence. 
\end{itemize}

The remainder of the article is organized as follows: Section III explains the proposed DT-native AI-driven service framework with the IoV use case implementation. The performance of the proposed model is evaluated in Section IV. Finally, Section V concludes the paper. 

\section{DT-Native AI-Driven System Architecture}
\subsection{Physical Twin Layer} 

\begin{figure*}[htbp]
\centerline{\includegraphics[width=0.9\linewidth]{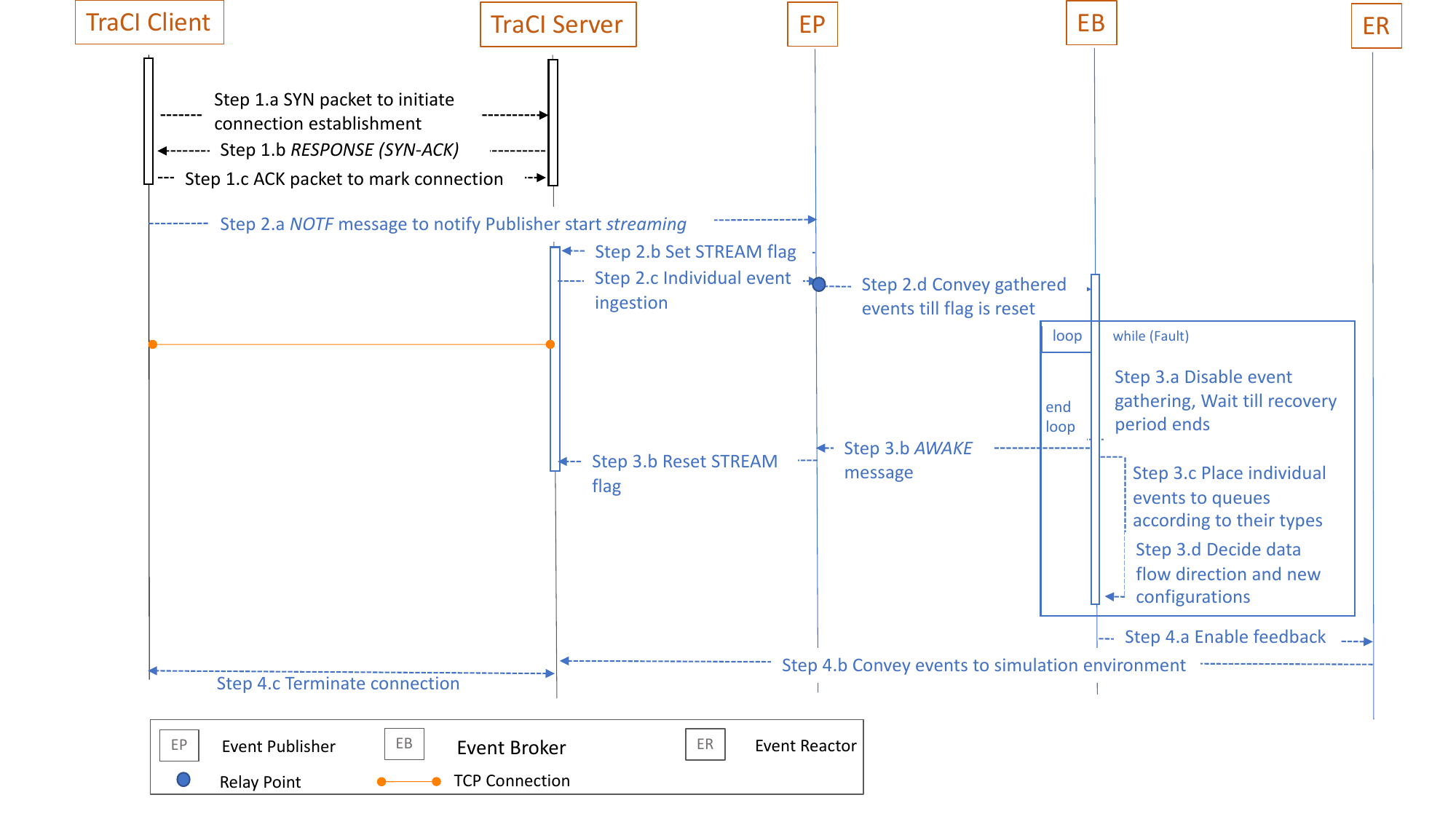}}
\caption{Proposed data flow pipeline between Physical Twin Layer and Digital Twin Layers.}
\label{two}
\end{figure*}

In this layer, we work on an Internet of Vehicles (IoV) scenario as a part of the industrial networks. This is because the connectivity problems between the sensors (traffic light sensors, buildings etc.), vehicles, and the infrastructure are becoming significant to maintain the needs of smart city applications. For this reason, we focus on a heterogeneous IoT environment that includes different types of sensors and moving vehicles environment. Therefore, this layer includes all physical entities within a transport network, such as lanes, road intersections, moving vehicles, IoT sensors, traffic lights etc. To construct a realistic traffic management scenario, we implement our Physical Twin Layer by using the microscopic and continuous traffic simulator Simulation of Urban MObility (SUMO). The block diagram of the proposed architecture is given in Fig. 1.

\subsection{Digital Twin Layer}
This layer includes the digital replica of physical sensors constructed by using the gathered event data. We deploy our Digital Twin models through a Docker Engine due to its efficiency in packaging and deploying different services with multiple clusters. Therefore, we deploy a Dockerfile with the base image and the dependencies to use as a data container in hosting the Extensible Markup Language (XML)-based twin sensor instances. In this way, we expect to improve the delivery of realistic insights as part of the Twin Service Layer operations. 

\begin{table} [!h]               
 \begin{center}
\caption{IoV Network Event Types\label{tab:table2}}
   \label{tab:table1}
    \begin{tabular}{l|l}
    \hline
\textit{Event} & \textit{Definition} \\
     \hline
     $E_1$ & (Vehicle ID, Location, Speed, Time Interval) \\
     $E_2$ & (Sensor ID, Location, Lane, Time Interval) \\
     $E_3$ & (Traffic light ID, Location, Lane, Duration Rate (red/green)\\
 \hline
    \end{tabular}
\end{center}
\end{table}

In the interaction between Digital Twin Layer and Physical Twin Layer, we define a novel data flow pipeline according to the needs of the simulation environment.
As we use the SUMO-based physical twin layer in our DT architecture, we implement Transmission Control Protocol (TCP) to initiate the TraCI interface of the SUMO. Thus, the pipeline entities, including the TraCI interface, are explained below:

\begin{itemize}
    \item \textit{TraCI Server:} TraCI permits the real-time streaming of implemented transport scenarios. Therefore, to enable interaction, a port on the physical road network side is calibrated for SUMO to act as a data server and start to listen to incoming connections.
    \item \textit{TraCI Client:} It is implemented in a Python script file to perform data retrieval requests.
    \item \textit{Event Publisher (EP):} After getting the successful TCP connection between the server and the client, the realistic road network data with junctions, vehicles, route directions, and speed information etc., are streamed by using the function list of TraCI. Thereafter, the gathered network information with the specified events is conveyed to the Event Broker. 
    \item \textit{Event Broker (EB):} It is used to transform the gathered events into individual virtual twins according to the event types. More specifically, we use a traditional event data (XML in our case) format. 
    \item \textit{Event Reactor (ER):} It passes the messages to the implemented Physical Twin Layer scenario. More specifically, it behaves as a feedback interface to reach the simulation environment and change the specified configurations. 
\end{itemize}

 The IoV-specific events from the experimental scenario are given in Table-I. As given in this table, we define three types of event classes in order to distinguish the transport network entities. In this circumstance, $E_1$ types imply the vehicle sensors with their location information (latitude and longitude), speed in $km/h$, and time interval in which the event is recorded. On the other side, $E_2$ type events stand for the IoT sensor recordings located within the transportation environment, throughout the roads and junctions. These environmental sensors include location information, lane information to describe their presence regarding the main roads, and time interval information. Lastly, $E_3$ type events stand for the traffic lights' recordings. These events include location information, lane information and additional duration rate of the red and green light signal timings. This attribute is used to decide the optimal number of vehicles in a road network to avoid congestion situations.

The data flow pipeline we proposed is given in Fig. 2. As seen from the figure; there are two main units, TraCI Client and TraCI Server, making the Physical Twin Layer and Digital Twin Layer connected by realizing TCP protocol. We denote these connection phases by marking them as the Step 1 category. More specifically, in step 1.a, client asks for a TCP connection by sending a $SYN$ packet to the server side. In step 1.b, the server responses with a $SYN-ACK$ packet to accept the connection request of the client. In step 1.c, the client sends an $ACK$ packet to the server and marks itself as connected. After the connection establishment, the client sends a $NOTF$ message to notify the EP to start data streaming. Then, the EP sets the $STREAM$ flag, and the server starts sending the data to the EP side. Here, EP behaves as a relay point by conveying the streamed data to the EB side. This event gathering to the EB side continues till the $STREAM$ flag is reset. In other words, the EB in our DT-native data flow pipeline gathers event information from EP as long as the TCP connection is established and the data streaming flag is set. Moreover, in case of any fault between these entities, $EB$ disables event streaming and waits till the recovery period ends. At the end of this period, it sends a $AWAKE$ message to $EP$ to trigger resetting the $STREAM$ flag. 
After that, $EB$ places the gathered events to separate queues to process according to their types. The event types are given in Table I, and explained in the above paragraph. The processing of event evaluations is performed by calculating their occurrence densities. For this, we calculate each of the event densities as,

\begin{equation}
\rho(E_i)=\frac{\sum_{k} O_i}{L_{Q_i}}\label{eq1}
\end{equation}

where \[1 \leq i \leq 3\ \text{,}\  1 \leq k \leq n\]

In this formula, $\rho(E_i)$ is the $i^{th}$ event density, $O_i$ is the occurrence of $i^{th}$ type events and $L_{Q_i}$ implies the number of events waiting in the related queue. In addition, $i$ is the number of event classes, and $k$ is the total streamed number of events. According to the calculated event densities, the $EB$ decides the data flow direction by considering the predefined threshold values. After deciding the data flow direction and the new configurations, the feedback interface is enabled to interact with the $ER$. By doing this, required changes are made to stabilize the realistic simulation scenario (step 4.b). After completing the data flow in both directions, the TCP connection is terminated between the client and server.

\subsection{Twin Service Layer}
This layer includes an RL engine to implement Reinforcement Learning algorithm to decide the optimal physical environment output recommendations to avoid congestion in an IoV scenario. More specifically, we have implemented the DDPG algorithm, which combines deep learning and policy gradient methods to learn optimal policies in the simulation environment with a continuous action space. As in our simulation environment, there are three types of events related to the vehicles, environment and traffic lights; the output of the DDPG algorithm will consist of three types of continuous actions, such as deciding a speed limit for vehicles flowing through a particular junction, and deciding the traffic lights' signalling time in order to prevent congestion scenarios. The block diagram for the proposed learner model within this layer is given in Fig. 3.

\begin{figure}[htbp]
\centerline{\includegraphics[width=1.1\linewidth]{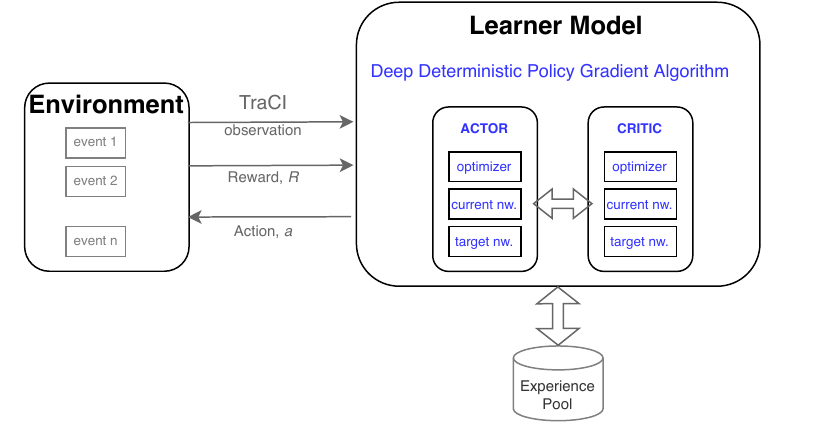}}
\caption{The interaction of environment and the learner model in implementing the DDPG algorithm.}
\label{three}
\end{figure}

Our learning model consists of four tuple \{$S_t$, $a$, $R$, $S_{t+1}$\}. Here $S_t$ is the current state and represents the total number of event classes in our IoV implementation scenario; $a$ is action space and decides the total number of sensor records minimizing the congestion probability in an IoV network; $R$ is an immediate reward, and $S_{t+1}$ is the decided next state. We apply DDPG algorithm by defining the reward function for state and action
pairs $(S_i, a_i)$. It executes the chosen action among a divergent set of actions and observes the next state and reward by storing the experience tuples in an experienced pool. As shown in Fig.3, the interaction between the simulation environment and the learner model is realized through the TraCI interface. The events are observed, and the actor and critic networks trace their change. 

The details of our proposed learner model consisting of an actor-network, a critic network and two target networks are explained below:

\begin{itemize}
\item \textit{Actor network 
 ($\theta^A$):} It learns the policy to select optimal actions. For example, in our IoV scenario, the actor-network tries to implement different speed limits to evaluate the results and decide on the optimal action. Also, this network can be called a policy network. 
    \item \textit{Critic network ($\theta^C$):} It learns to evaluate the quality of the chosen actions. For instance, it evaluates if a decided speed limit for a vehicle cause to any congestion, resulting in our IoV scenario implementation. Also, this network can be called a value function network.
    \item \textit{Actor target network ($\theta^{A^{'}}$):} It is periodically updated during the learning process. It is a time-delayed copy of the actor-network.
    \item \textit{Actor critic network ($\theta^{C^{'}}$):} It is periodically updated during the learning process. It is a time-delayed copy of the critic network. 
    \end{itemize}

The DDPG algorithm updates the Q values based on the environment and learning rate values by using the Bellman equation: 

\begin{equation}
y_i= R_i + \alpha Q^{'} + (S_{i+1},C^{'}(S_{i+1}|\theta^{C^{'}})|\theta^{Q^{'}})\label{eq}
\end{equation}

In this equation, $\alpha$ is the learning rate and takes values from a predefined interval. We also try to minimize the mean squared loss between the updated Q value and the originally calculated Q value given as:

\begin{equation}
Loss= \frac{1}{E} \sum_{i} (y_i-A(S_i, a_i|\theta^{A}))^2\label{eq2}
\end{equation}

In this equation, $E$ implies the total number of performed experiences during the algorithm run with the defined optimizer function. 

    
\section{Performance Evaluation}

In this section, we show the performance of our proposed DT-native service architecture by considering (i) the efficiency of the proposed data flow pipeline by changing the number of sensors in the implemented IoV scenario and observing the average processing time and (ii) the efficiency of the proposed learner model by measuring its error and the resultant congestion scenarios. The simulation parameters are given in Table II. 

\textit{Experimental Setup:} We carried out Physical Twin Layer implementations on SUMO\textsuperscript{\copyright} environment. We deployed the Digital Twin layer over a Docker container. Also, as the basis of our learning model implementation, we used the TensorFlow library.

\begin{table}[thpb!]
    \centering
\caption{Simulation Parameters} 
\centering 
    \begin{tabular}{l c} 
\hline 
Parameters & Values \\ [0.5ex]
\hline\hline 
Number of event classes &  \{1, 2, 3\} \\ 
Total number of sensors & \{5, 20, 90\}\\
Learning rate for policy networks & [0.001, 0.2]  \\
Learning rate for Q-networks & [0.001, 0.2]  \\
Optimizer & adam  \\
Discount factor  & 0.8 \\
Batch size & \ 256\\
Confidence interval  & 95\% \\
\hline 
    \end{tabular}
\label{tab:hresult}
\end{table}

\begin{figure}[htbp]
\centerline{\includegraphics[width=\linewidth]{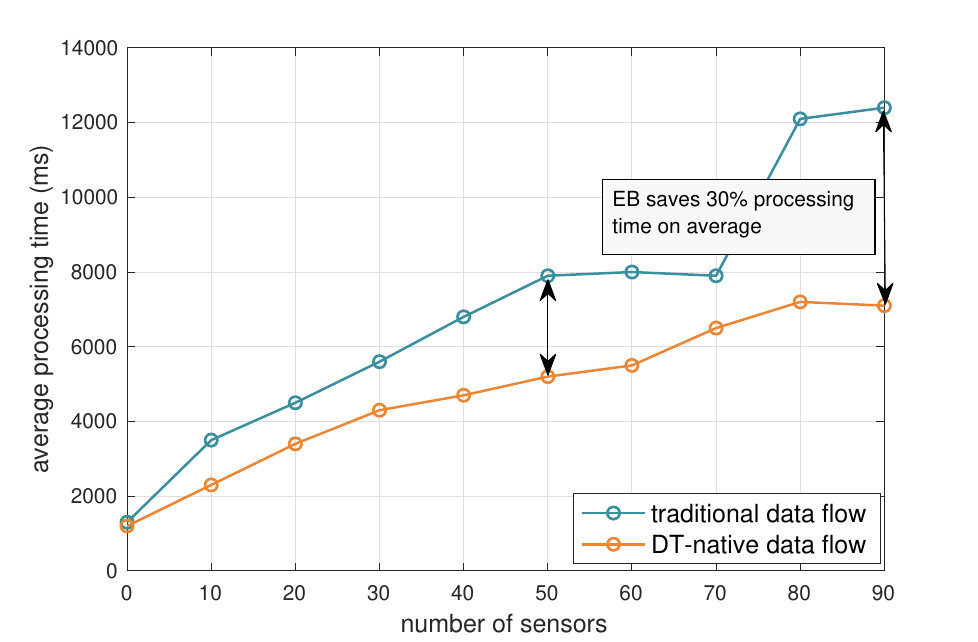}}
\caption{Average processing time for traditional data flow and proposed DT-native data flow.}
\label{four}
\end{figure}

To measure the performance of the proposed DT-native data flow architecture, we consider three types of events collected for the increasing number of sensors. For this, we assume there are five IoT sensors in the initial state of the simulation. After performing the event collection as described in the above sections, we increased the number of sensors to twenty and then to ninety. We perform this increment in order to observe the behaviour of entities, such as EP, EB, and ER, which perform between the Physical Twin Layer and Digital Twin Layers of our DT model. We observe the average processing time of the proposed DT-native data flow and compare it with the manual data flow trigger approach as given in \cite{t6conf}. Our recorded results are given in Fig. 4. We notice that the processing time is saved by ~30\% on average with the proposed DT-native data flow. This is because the EB keeps track of the specified events and calculates density values for each of these event types. According to the calculated values, it triggers the flow direction and applies the optimal number of sensors in order to avoid congestion scenarios in an IoV scenario. Thanks to this cognitive approach, the proposed pipeline can rapidly handle dynamic scenarios. 

\begin{figure}[htbp]
\centerline{\includegraphics[width=0.9\linewidth]{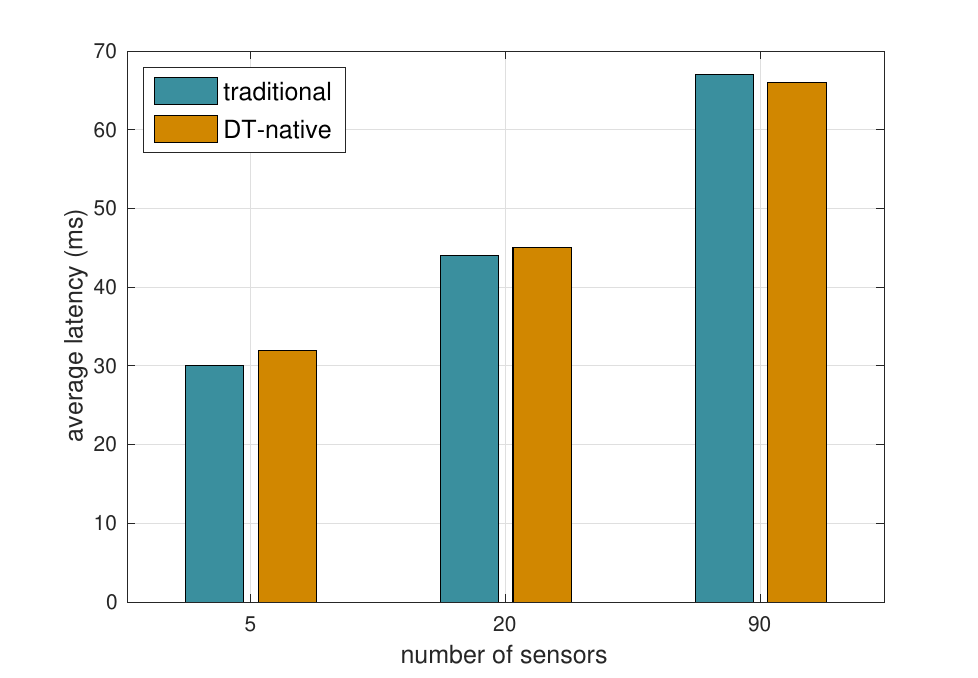}}
\caption{Average latency for traditional data flow and proposed DT-native data flow.}
\label{five}
\end{figure}

Moreover, we investigate the efficiency of the proposed DT-native data flow architecture by measuring the latency values for the implemented simulation run, which is realized in the former step. We record the latency values in $ms$ for five, twenty, and ninety sensor cases. As seen in Fig. 5, the traditional and DT-native data flow pipelines have similar latency values for each scenario. The main reason is that the EB performs event analysis and density calculation in DT-native data flow. This step directly causes operational latency in the data streaming process. Therefore, if an IoT network consists of a high amount of sensors, the proposed DT-native data flow pipeline performs favourably in terms of average latency when compared to the case when we implement ninety sensors. 
\begin{figure}[htbp]
\centerline{\includegraphics[width=\linewidth]{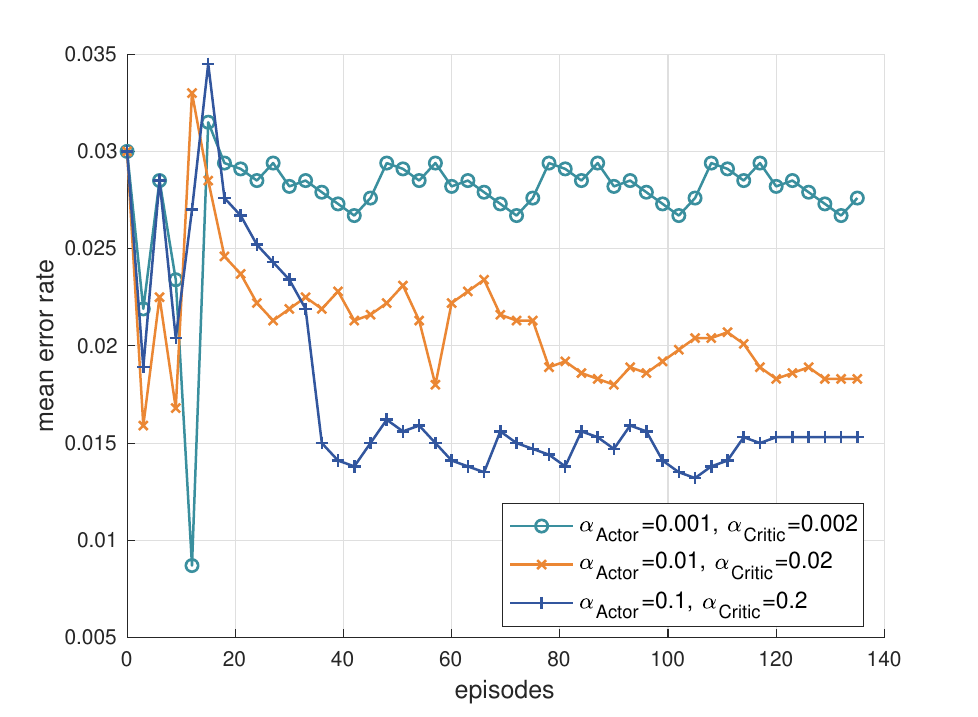}}
\caption{The success of the DDPG algorithm with changing learning rate combinations for actor and critic networks.}
\label{six}
\end{figure}

Furthermore, to measure the performance of the proposed learner model, we perform DDPG with the changing combinations of learning rates considering the interval $[0.001, 0.2]$ for actor and critic networks. We assume that optimizers in actor and critic networks use the same optimization function, $adam$. Also, throughout all the episodes run, we set the discount factor value to 0.8 to help the algorithm converges. Moreover, we use a batch size of 256. We have used the TensorFlow library in order to run the DDPG algorithm. Our recorded results show three learning rates combination for actor and critic networks for 135 performed episodes. As seen from Fig. 6, the most successive learning rates combination, ($\alpha_{Actor}$, $\alpha_{Critic}$), is observed as (0.1, 0.2) with the less mean error rates. In addition, the case with the learning rates (0.01, 0.02) also converges at the end of the realized episodes as compared. On the contrary, the learning rates combination (0.001, 0.002), shown as a light blue line in Fig. 6, does not converge through all performed episodes. This means that learning rates set in this case are insufficient for the learner model to cover all event occurrences in the simulation. We argue that the learning rates should be set above these values to develop an effective DDPG run.

\section{Conclusion}
In this study, we focus on three distinct challenges of IoT networks; perpetual connectivity, sensor data streaming, and depth of the data insights. Considering the need for continuous monitoring and enhanced analysis of IoT networks, we propose a DT-native AI-driven service framework with a TCP-based data flow pipeline and a Reinforcement Learning (RL)-based learner model. We apply our proposed framework to one of the broad concepts of IoT networks, the Internet of Vehicles (IoV) network scenario, and measure its efficiency for processing time and latency. Our simulation results show ~30\% processing time saving with the proposed TCP-based data flow pipeline as compared to manual data flow triggering. Furthermore, we measure the efficiency of our proposed learner model with several learning rates combination and come up with the most successive version for actor and critic networks.


\bibliographystyle{IEEEtran}
\bibliography{IEEEabrv,references}

\vspace{12pt}

\end{document}